\documentclass[fleqn,twoside]{article}
\usepackage{espcrc2}

\usepackage{graphicx, epsfig}
\usepackage[figuresright]{rotating}

\newcommand{\AmS}{{\protect\the\textfont2
  A\kern-.1667em\lower.5ex\hbox{M}\kern-.125emS}}

\title{Quark mass dependence of pseudo-scalar masses and coupling constants}

\author{qq+q Collaboration \\[0.5em]
	F. Farchioni\address[Muenster]{Institut f\"ur Theoretische Physik, 
	Universit\"at M\"unster, Wilhelm-Klemm-Str. 9,\\ 
	D-48149 M\"unster, Germany},
	C. Gebert\address[desy]{
        Deutsches Elektronen-Synchrotron DESY,
        Notkestr.\,85, D-22603 Hamburg, Germany},
	I. Montvay\addressmark[desy], E. Scholz\addressmark[desy], 
	L. Scorzato\addressmark[desy].}
       
\begin{document}

\begin{abstract}
The dependence of pseudo-scalar masses and decay constants on the sea and 
valence quark masses is investigated in the pseudo-Goldstone boson sector 
of QCD with two light quark flavours. The sea quark masses are at present 
in the range $\frac{1}{3}m_s \leq m_{ud} \leq \frac{2}{3}m_s$ whereas the 
valence quark masses satisfy $\frac{1}{2}m_{sea} \leq m_{val} \leq 2m_{sea}$. 
The values of the Gasser-Leutwyler low energy constants $L_4$, $L_5$, $L_6$ 
and $L_8$ are estimated. The computation is done with the Wilson-quark lattice 
action at gauge coupling $\beta=5.1$ on $16^4$ lattices. 
${\cal O}(a)$ effects are taken into account by applying chiral perturbation 
theory for the Wilson lattice action as proposed by Rupak and Shoresh.
\vspace{1pc}
\end{abstract}

\maketitle

\section{Introduction}
Monte Carlo simulations of QCD with dynamical quarks are
done in most cases at relatively large quark masses (typically
two quark flavours with $m_{ud} > m_s/2$).
This makes the extrapolation to the physical point 
$m_{ud} \simeq  m_s/20$
rather uncertain.
The extrapolation is done by using (PQ)ChPT -- typically
to NLO (1-loop) order.
Estimates \cite{SHARPE-SHORESH,DURR} show that one should
perform simulations in the range 
$m_{ud} < m_s/4, m_s/3$ 
in order to see the expected logarithmic dependence.
Matching the predicted functional dependence is a crucial
test for Lattice QCD. 

Until recently the comparison between lattice data and ChPT was not
satisfactory. In a recent paper \cite{qqq51} we suggested that the agreement
can be found when going to light enough dynamical quarks. Here we supplement some 
integration of the analysis of \cite{qqq51} and provide some further comments.

\section{Simulations}
For our simulation we used the algorithm described in 
\cite{TSMB} and further improved as in 
\cite{qqq51} and references therein.
In our range of parameters we found that the cost for
producing one independent gauge field configuration roughly goes as:
\[
\mbox{Cost} \simeq F  \,  (a m_q)^{-2} \, \Omega , 
\]  
where $am_q$ is the quark mass in lattice unit, $\Omega$ is the number of lattice points
and the factor $F\simeq 10^7$flop for the plaquette, but one order of magnitude smaller for 
$m_{\pi}$ and $f_{\pi}$.

We produced three sets of $O(1000)$ thermalized configurations for $N_f=2$ 
unimproved Wilson fermions at  Vol$=16^4$,  $\beta=5.1$ and $\kappa=0.176, 0.1765, 0.177$.
From these points we extrapolated a value of $r_0/a$ at $\kappa_{cr}=0.1773(1)$ which is
$r_0(\kappa_{cr})/a=2.65(7)$. This corresponds
to an UV cutoff $a=0.189(5)$fm $\simeq (1.04 \mbox{ GeV})^{-1}$, and to a physical volume
 $L\simeq 3$fm. In those three points we found a ratio $m_q/m_s$ equal
to $0.68$, $0.59$ and $0.35$ respectively. Here $m_q$ represents the sea quark mass
defined as $M_r=(m_{\pi} r_0)^2$, when the strange quark mass  corresponds to
$M_r=3.1$ \cite{NF2TEST}.
For the lightest valence quarks the ratios $m_q^{val}/m_s$  
become respectively $0.37$, $0.34$ and $0.17$. 
Even with these small quark masses finite volume effects are expected to be under control
since we have always $L m_{\pi}^{val} \geq 4.8$. Of course this is payed with a very low
UV cutoff, and one expects large lattice artifacts. These are taken into account in the
analysis.

\begin{figure}[htb]
\vspace{-0.5em}
\begin{center}
\epsfig{file=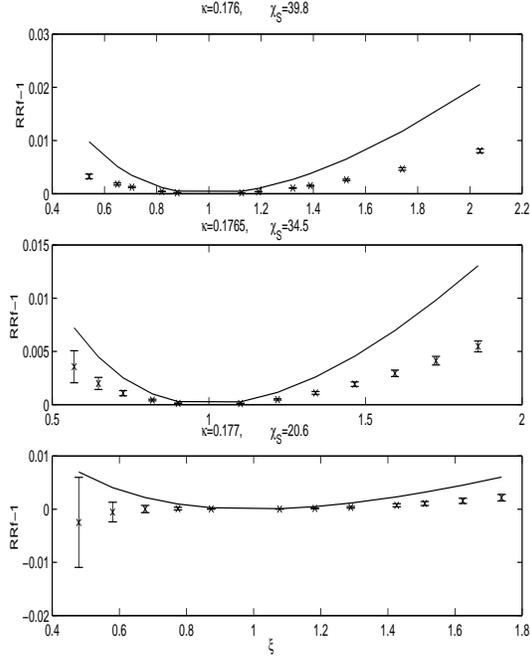,width=7cm,height=9cm,angle=0}
\end{center}
\vspace{-3.0em}
\caption{\label{fig:RRf}
$RRf-1$ predictions against the data.
}
\vspace{-0.5em}
\end{figure}
\begin{figure}[htb]
\vspace{-0.5em}
\begin{center}
\epsfig{file=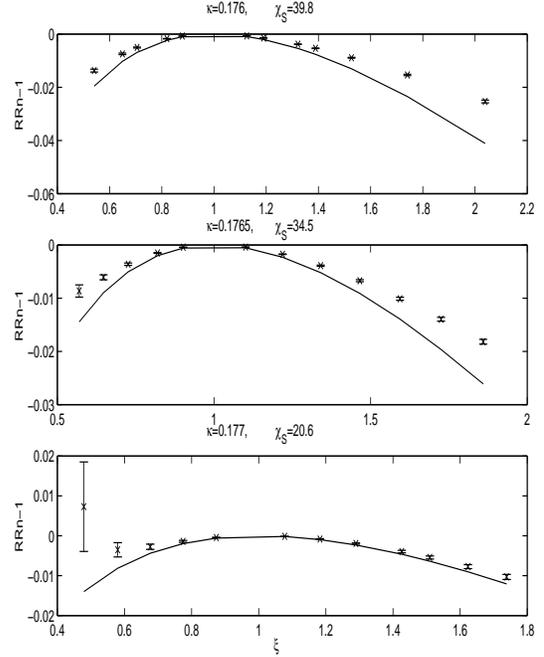,width=7cm,height=9cm,angle=0}
\end{center}
\vspace{-3.0em}
\caption{\label{fig:RRn}
$RRn-1$ predictions against the data.
}
\vspace{-0.5em}
\end{figure}

\section{Comparison with ChPT}

Our first goal is to confront the results of numerical simulations with the (PQ)ChPT 
formulas \cite{BERNARD-GOLT,SHARPE-SHORESH}. In order to cancel the $Z$-factors of multiplicative
renormalization, which in the case of a mass-independent renormalization scheme only
depend on the gauge coupling and not on the quark mass, we considered ratios of quark masses
($m_{qA}$), pion masses ($m_{AB}$) and pion decay constants ($f_{AB}$).
Here $A,B$ stand for the flavor indices of valence ($V$) or sea ($S$) quarks. 

If we assume that there are no lattice artifacts, no NNLO corrections
and we take for $\chi_S=\frac{2 B_0 m_{qS}}{f_0^2}$ its tree level estimate, 
$\chi_S^{est} \equiv  \frac{(r_0 m_{\pi})^2}{(r_0 f_0)^2}$,
then the ratios (here $\xi \equiv  \frac{\chi_V}{\chi_S}$)
\begin{eqnarray*}
RRf \hspace{-7pt} &\equiv & \hspace{-7pt} \frac{f_{VS}^2}{f_{VV}f_{SS}} =
1 + \frac{\chi_S^{est}}{32N_s\pi^2}[\xi-1 -\log( \xi)],  \\
RRn \hspace{-7pt} &\equiv & \hspace{-7pt}
\frac{4\xi m_{VS}^4}{(\xi\hspace{-3pt}+\hspace{-3pt}1)^2m_{VV}^2 m_{SS}^2}
\hspace{-3pt}= \hspace{-3pt}1 + \hspace{-3pt}\frac{\chi_S^{est}
[\log(\xi)\hspace{-3pt}-\hspace{-3pt}\xi \hspace{-3pt}+\hspace{-3pt}1]}{16N_s\pi^2} 
\end{eqnarray*}
are non trivial and parameterless predictions of ChPT. 
This provides a strong check of how far we are from the NLO ChPT regime.
In figure \ref{fig:RRf} and \ref{fig:RRn} we superimpose the predicted functions
to the data. The agreement is of course not complete 
(there are indeed $O(a)$ and NNLO effects), but the corrections are sub-dominant
contributions. Moreover the agreement improves when the sea quark masses decrease.

\begin{figure}[htb]
\vspace{-0.5em}
\begin{center}
\includegraphics[width=7cm,height=7cm,angle=-90]{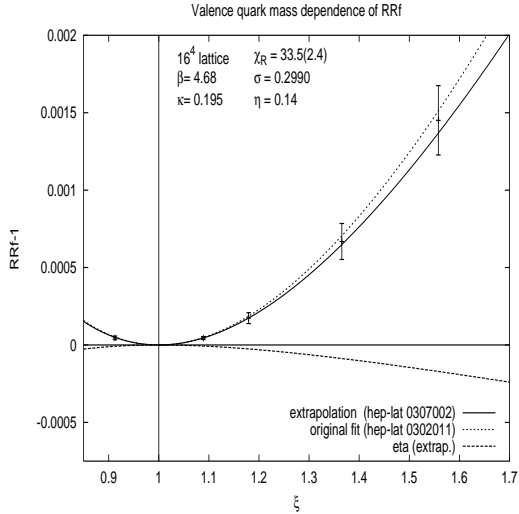}
\end{center}
\vspace{-3.0em}
\caption{\label{fig:scarr}
Comparison of the extrapolation of the fit of $RRf$ from $\beta=5.1$ \cite{qqq51} 
with the original fit.
}
\vspace{-0.5em}
\end{figure}
\begin{figure}[htb]
\vspace{-0.5em}
\begin{center}
\includegraphics[width=7cm,height=7cm,angle=-90]{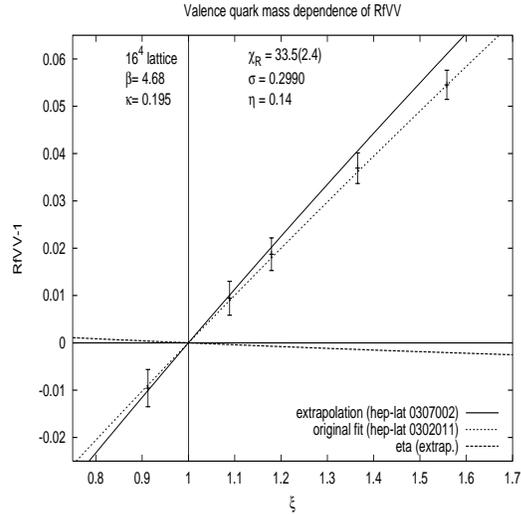}
\end{center}
\vspace{-3.0em}
\caption{\label{fig:scafvv}
Comparison of the extrapolation of the fit of $Rf_{VV}$ from $\beta=5.1$ \cite{qqq51} 
with the original fit.
}
\vspace{-0.5em}
\end{figure}

Encouraged by this results, we systematically compared
our data with those ratios of pion masses and coupling constants
which determine the Gasser-Leutwyler \cite{CHPT} coefficients
$L_4$, $L_5$, $L_6$ and $L_8$. Since we expected
large lattice artifacts in the data, the comparison was done
with W(PQ)ChPT \cite{RUPAK-SHORESH}, including $O(a)$ lattice
artifacts in the effective continuum theory. Besides that,
we also included the relevant contributions from 
NNLO ChPT \cite{SHARPE-VDWATER}. Although
this involve many parameters, one can obtain enough constraints
from Partially Quenched simulations. The details of the fitting
procedure and the results are described in \cite{qqq51}.
Here we simply add that, following the analysis of 
\cite{BARUSH}, we can now include also $O(a^2)$ lattice
artifacts in our fit of the pion mass ratios. It turns
out that this does not add new parameters to the fit, but it
amounts to a redefinition of the Wilson-ChPT coefficients $W_i$.

In order to have a consistency check of the surprisingly
small lattice artifacts that we found, we compared the results
of \cite{qqq51} with those obtained in an older simulation with larger
lattice spacing ($\beta =4.68$, $a\simeq 0.27$fm) \cite{VALENCE}.
The comparison (fig. \ref{fig:scarr} and 
\ref{fig:scafvv}) shows quite small scale breaking. Larger scale breaking ($\sim 8\%$)
are observed for the ratio $m_{\pi}/m_{\rho}$, which at fixed $M_r=1.088$ goes
from $0.613$ (at $\beta=5.1$) to $0.664$ (at $\beta=4.68$). 

To summarize: we showed that it is possible to simulate, 
with reasonable costs, very light dynamical
quarks. Compensating $O(a)$ effects by introducing $O(a)$ terms in the PQCh-Lagrangian is,
in this case, a viable alternative to $O(a)$ improvement of the action.
The observed $O(a)$ contributions in the light Goldstone boson sector are surprisingly small,
while NNLO are still important.
The expected behavior predicted by PQChPT is already visible, although a quantitative 
determination of the LEC's still needs further simulations 
at smaller masses and lattice spacing.
Most of the numerical calculations presented here have been done at the
computers of NIC-Juelich and Zeuthen.

\end{document}